\documentclass{pramana}
\usepackage[nosort]{cite}

\usepackage{graphicx,amsmath,bm}
\usepackage{flushend}
\usepackage{lipsum}
\usepackage{dingbat}
\usepackage[vcentermath]{youngtab}
\usepackage{subfigure}
\usepackage{multicol}
\usepackage{multirow}
\usepackage{amsmath,cuted}
\usepackage{epsfig}
\usepackage{graphicx}
\usepackage[all,knot]{xy}
\xyoption{arc}
\usepackage{hyperref}
\usepackage{times}
\usepackage{graphics}
\usepackage{amsfonts}
\usepackage{wrapfig}
\usepackage{color, xcolor}
\usepackage{verbatim}
\usepackage{graphics}
\newcommand{\beq}{\begin{equation}}
\newcommand{\eeq}{\end{equation}}
\newcommand{\bea}{\begin{eqnarray}}
\newcommand{\eea}{\end{eqnarray}}
\newcommand{\bit}{\begin{itemize}}
\newcommand{\eit}{\end{itemize}}
\newcommand{\bmt}{\begin{pmatrix}}
\newcommand{\emt}{\end{pmatrix}}
\newcommand{\btab}{\begin{tabular}}
\newcommand{\etab}{\end{tabular}}

\newcommand{\bframe}{\begin{frame}}
\newcommand{\eframe}{\end{frame}}
\newcommand{\bblock}{\begin{block}}
\newcommand{\eblock}{\end{block}}

\newcommand{\nnu}{\nonumber\\}

\newcommand{\oot}{\overline {126}}
\newcommand{\boot}{${\bf{\oot}}$ }

\newcommand{\nn}{\nonumber\\}
\renewcommand{\arraystretch}{2.0}
\newcommand{\ebar}{\bar{e}}
\newcommand{\dbar}{\bar{d}}
\newcommand{\ubar}{\bar{u}}
\newcommand{\hbaar}{\bar{h}}
\newcommand{\nubar}{\bar{\nu}}
\newcommand{\nbar}{\bar{n}}
\newcommand{\onbb}{0\nu\beta\beta}

\newcommand{\Abar}{\bar{A}}

\newcommand{\Cbar}{\bar{C}}
\newcommand{\Dbar}{\bar{D}}
\newcommand{\Ebar}{\bar{E}}
\newcommand{\Fbar}{\bar{F}}

\newcommand{\Jbar}{\bar{J}}
\newcommand{\Kbar}{\bar{K}}
\newcommand{\Lbar}{\bar{L}}

\newcommand{\Obar}{\bar{O}}
\newcommand{\Pbar}{\bar{P}}

\newcommand{\tbar}{\bar{t}}

\newcommand{\Wbar}{\bar{W}}

\newcommand{\bB}{{\color[HTML]{0000DD} B}}
\newcommand{\bI}{{\color[HTML]{0000DD} I}}
\newcommand{\bQ}{{\color[HTML]{0000DD} Q}}
\newcommand{\bR}{{\color[HTML]{0000DD} R}}
\newcommand{\bS}{{\color[HTML]{0000DD} S}}
\newcommand{\bU}{{\color[HTML]{0000DD} U}}
\newcommand{\bV}{{\color[HTML]{0000DD} V}}
\newcommand{\bX}{{\color[HTML]{0000DD} X}}
\newcommand{\bY}{{\color[HTML]{0000DD} Y}}
\newcommand{\bZ}{{\color[HTML]{0000DD} Z}}
\newcommand{\blue}{{\color[HTML]{0000DD} Blue\,}}

\newcommand{\rA}{{\color[HTML]{DD0000} A}}
\newcommand{\rC}{{\color[HTML]{DD0000} C}}
\newcommand{\rD}{{\color[HTML]{DD0000} D}}
\newcommand{\rE}{{\color[HTML]{DD0000} E}}
\newcommand{\rF}{{\color[HTML]{DD0000} F}}
\newcommand{\rG}{{\color[HTML]{DD0000} G}}
\newcommand{\rh}{{\color[HTML]{DD0000} h}}
\newcommand{\rJ}{{\color[HTML]{DD0000} J}}
\newcommand{\rK}{{\color[HTML]{DD0000} K}}
\newcommand{\rL}{{\color[HTML]{DD0000} L}}
\newcommand{\rM}{{\color[HTML]{DD0000} M}}
\newcommand{\rN}{{\color[HTML]{DD0000} N}}
\newcommand{\rO}{{\color[HTML]{DD0000} O}}
\newcommand{\rP}{{\color[HTML]{DD0000} P}}
\newcommand{\rS}{{\color[HTML]{DD0000} S}}
\newcommand{\rt}{{\color[HTML]{DD0000} t}}
\newcommand{\rW}{{\color[HTML]{DD0000} W}}

\newcommand{\rAbar}{{\color[HTML]{DD0000} \bar{A}}}
\newcommand{\rCbar}{{\color[HTML]{DD0000} \bar{C}}}
\newcommand{\rDbar}{{\color[HTML]{DD0000} \bar{D}}}
\newcommand{\rEbar}{{\color[HTML]{DD0000} \bar{E}}}
\newcommand{\rFbar}{{\color[HTML]{DD0000} \bar{F}}}

\newcommand{\rhbar}{{\color[HTML]{DD0000} \bar{h}}}
\newcommand{\rJbar}{{\color[HTML]{DD0000} \bar{J}}}
\newcommand{\rKbar}{{\color[HTML]{DD0000} \bar{K}}}
\newcommand{\rLbar}{{\color[HTML]{DD0000} \bar{L}}}

\newcommand{\rObar}{{\color[HTML]{DD0000} \bar{O}}}
\newcommand{\rPbar}{{\color[HTML]{DD0000} \bar{P}}}

\newcommand{\rtbar}{{\color[HTML]{DD0000} \bar{t}}}
\newcommand{\rWbar}{{\color[HTML]{DD0000} \bar{W}}}
\newcommand{\rnubar}{{\color[HTML]{DD0000} \bar{\nu}}}
\newcommand{\red}{{\color[HTML]{DD0000} Red\,}}

\newcommand{\gubar}{{\color[HTML]{00AA11} \bar{u}}}
\newcommand{\gdbar}{{\color[HTML]{00AA11} \bar{d}}}
\newcommand{\gQ}{{\color[HTML]{00AA11} Q}}
\newcommand{\gebar}{{\color[HTML]{00AA11} \bar{e}}}
\newcommand{\gL}{{\color[HTML]{00AA11} L}}
\newcommand{\gH}{{\color[HTML]{00AA11} H}}
\newcommand{\gHbar}{{\color[HTML]{00AA11} \bar{H}}}
\newcommand{\green}{{\color[HTML]{00AA11} Green\,}}

\begin{document}
\onecolumn

\title{Effective Sextic Superpotential and $B-L$ violation in NMSGUT}

\author{C. S. Aulakh, R. L. Awasthi$^{*}$ \and Shri Krishna}
\affilOne{Indian Institute of Science Education and Research Mohali\\}

\twocolumn[{

\maketitle

\hspace{1.3cm} $^*$Speaker: awasthi@iisermohali.ac.in \\
%%include \corres to print the corresponding author Email id
%\corres{awasthi@iisermohali.ac.in}

\begin{abstract}
We list operators of the superpotential of the  effective MSSM
that emerges from the NMSGUT up to sextic degree. We give
  illustrative expressions for the coefficients in terms
of NMSGUT parameters. We also estimate the impact of GUT scale
threshold corrections on these effective operators in view of the
demonstration that $B$ violation via quartic superpotential terms
can be suppressed to acceptable levels after including such
corrections in the NMSGUT. We find a novel $B, B-L$ violating
quintic operator that leads to the decay mode $n\to e^- K^+$. We
also remark that the threshold corrections to the Type I seesaw
mechanism make the deviation of right handed neutrino masses from
the GUT scale more natural while  Type II seesaw neutrino masses,
which   earlier tended to utterly  negligible  receive threshold
enhancement. Our results are of relevance for analyzing $B-L$
violating operator based, sphaleron safe, Baryogenesis.

\end{abstract}

%%insert keywords separated by comma using \keywords{words}

%%include \pacs{number} to print the PACS number

}]

%\doinum{12.3456/s78910-011-012-3}
%\artcitid{\#\#\#\#}
%\volnum{123}
%\year{2016}
%\pgrange{23--25}
\setcounter{page}{1}
\lp{9}

\section{Introduction}

Despite various open questions of principle, such as a mechanism to ensure
light Standard Model (SM) Higgs doublets, and still without any direct
experimental proof, particularly proton decay, Grand Unification
models based on SO(10) remain the most transparent framework to think
about beyond standard model (BSM) physics. In such grand unification
theories (GUTs)  neutrino masses are tightly connected to rest of the fermion
masses of the SM and other high scale parameters. They explain charge quantization,
give gauge and matter (quark- lepton) unification and predict important exotic
processes (proton decay) and parameters (neutrino masses and mixing).

The effective field theory (EFT) of non- renormalizable
higher dimensional operators (HDO) formed from the fields of the
SM plays an essential role in BSM studies. While  the
bottom-up  approach to the effective theory  makes fewer
speculations it also gives no clue regarding the scale of new
physics and the sizes of  the coupling coefficients of HDOs; or
rather it fails to utilize the few but strong hints about BSM
scales available from the size of neutrino masses and the
unification of couplings.  However given a viable GUT, specially a
supersymmetric (SUSY) one, it is straightforward to derive the HDOs that
correct the leading order renormalizable (minimal supersymmetric) 
standard model, (MS)SM,  that arises from
the GUT when superheavy fields are set to zero, simply by solving
the (algebraic) equations for the heavy (super)fields in the
approximation that their momenta are negligible compared to their
masses. Exotic operators that violate accurate but apparently
accidental symmetries of Standard Model, such as Baryon ($B$) and
Lepton ($L$) numbers  are obviously of most interest for developing
expectations regarding the experimental implication of any
particular UV completion of the SM. HDOs  which break $B$ and $L$
symmetry while preserving $B-L$ are familiar consequences of GUTs.
On the other hand Majorana neutrino masses $(\Delta L \neq 0)$
require breaking of $B-L$ symmetry  and in fact arise in
renormalizable models of Type I and Type II seesaw via vacuum expectation
values (VEVs) of $B-L$ non-singlet fields.  $B-L$ violating processes  may be
particularly important for GUT scale baryogenesis because a baryon
asymmetry produced via~$B-L$ conserving operators is liable to
washout via Electroweak sphaleron processes that are unsuppressed
above $M_W$. On the other hand baryon asymmetry  generated via
$B-L$ violating operators that also violate $B$ can generate a
Baryon asymmetry that survives Spahleron washout\cite{babumohap}.
This  revives the possibility that the observed baryon asymmetry
may arise at GUT scales rather than by sphaleron reprocessing of
the   Lepton asymmetry generated via $B-L$ violating  decay of
heavy righthanded (RH) neutrinos \cite{leptogenesis}. It is interesting that this
possibility is proposed in the context of the same theories with
gauged $B-L$ (such as SO(10) GUTs) that provide a natural context
for Type I and Type II Seesaw masses.

Minimal Supersymmetric SO(10) Grand Unification Theory
(MSGUT)\cite{aulmoh,ckn,abmsv} has been
developed over a long period into a   completely realistic scenario compatible with
known processes, data and structures of physics up to several hundred GeV, including
the critical BSM phenomenon of neutrino masses, to an extent
that is not rivaled by any other model. Moreover the solubility of the
spontaneous symmetry breaking (SSB) at GUT scales  permits expression of SM
data in terms of GUT couplings,
explicit  evaluation of GUT scale threshold corrections and determination of
viable $B$ violation  operators. There are  a number of  other  attractions of the
New-MSGUT (NMSGUT): Minimal number of parameters, R-parity preserving effective MSSM giving
viable SUSY WIMP Dark Matter, large soft trilinear ($A_0$) indications in advance
of Higgs discovery\cite{nmsgut} etc. Thus the evaluation of the exotic operators
for this theory is of relevance. An attempt  in this direction for a closely related
model which modifies the NMSGUT just  by adding an additional $\bm{10}$-plet
(and thus shares the same SSB structure at high scales) has already
appeared\cite{nathraza}. It uses the rather unwieldy decomposition
of SO(10) invariants via $\rm SU(5)\times U(1)$ maximal subgroup of SO(10). However the
NMSGUT  analysis mentioned above was via the more symmetric, and thus somewhat
more convenient and transparent,  decomposition of SO(10) in terms
of the Pati-Salam (PS) group $ G_{PS}\equiv \rm SU(2)_L\times
SU(2)_R\times SU(4)$. Thus an evaluation of the effective
operators and their coefficients in terms of the GUT coupling
using this method is both possible (given the detailed
decompositions we generated   during  our previous
calculations\cite{ag1,ag2,nmsgut,bstabhedge}) and required
considering the need to evaluate the workability of the  NMSGUT as
a comprehensive unification of particle physics i.e.  beyond just
fitting of SM data and compatibility with exclusion limits on
exotic processes. In this proceeding we report on our calculation
of the the effective operators possible in the
NMSGUT superpotential  up to terms sextic in the chiral fields and
on their coefficients in terms of GUT parameters up to terms quintic.
We also found one new operator which was missing in the literature which is
relevant to the study of nucleon decay process in $B-L$ violating
baryogenesis.

Techniques for computing the decomposition of \hfil\break SO(10)
invariants in terms of PS along with coefficients of
dimension 5 operators for $B$, $L$ violation were presented in
\cite{ag1, nmsgut}. The complete spectrum was presented in
\cite{ag2} for MSGUT and in \cite{nmsgut} for NMSGUT.
See\cite{bmsv,fuku04} for related spectrum calculations.
Decomposition of the NMSGUT invariants which contain Higgs
doublets  as one of the fields in the invariants were evaluated
for the computations in \cite{bstabhedge} and are thus available
to us.

  In Section~\ref{sec:nmsgut} we discuss  the effective MSSM theory
emerging from NMSGUT. In Section~\ref{sec:threnh} we will
recapitulate the impact of GUT scale threshold corrections on the
strength of coefficients of operators and their impact on proton
lifetime estimation. In Section~\ref{sec:effbl} we discuss impact
of threshold correction on $B-L$ violating processes.
%In Section~\ref{sec:baryo} we briefly discuss the origin of baryon
%asymmetry of universe (BAU) through GUT baryogenesis.
In Section~\ref{sec:conclusion} we conclude. Tables of effective
superpotential operators    are provided    in the  Appendix.

\section{Effective MSSM from NMSGUT}\label{sec:nmsgut}

Effective theories derived by integrating out heavy fields are
familiar from the basic example of the Fermi theory of beta decay
that arises from the Electro-weak theory by integrating out the
heavy  $W,Z$ gauge bosons. The most familiar paradigm for the
effective  Lagrangian  due to integrating out heavy matter fields
is Weinberg's \cite{Weinberg:1979sa} unique, $L$ and $B-L$
violating, dimension five,  Majorana mass  term for left handed
neutrinos. It arises from integrating out right handed neutrinos
with  large ($B-L$ violating) Majorana masses  : $ m_{\nu_L} \sim
Y^{\nu}_{AB}{L_A H L_B H}/ M_{\nu_R}$.  If the neutrino Yukawa
couplings $Y^{\nu}_{AB} $ are similar in size to the Yukawa
couplings for charged fermions of their generation, as expected
from GUTs, specially SO(10) GUTs,  then the required scale of new
physics  indicated by the measured neutrino masses is near the GUT
scale. The scale of new physics can be lowered if these couplings
are much smaller. If the leading order contributions are
completely absent, or when violation of superselection rules
applicable to the lowest dimension effective operators is
relevant (as for $B$ violation or when $B-L$ violating operators
enable sphaleron-safe baryogenesis)  higher dimensional operators
may be significant and must be considered.

Neutrino masses and mixing explained through seesaw mechanism
assume a Majorana nature for neutrinos. On the other hand it is
still perfectly feasible that right handed neutrinos are in fact
light particles and that neutrino masses are of Dirac type.
Oscillation experiments cannot distinguish between these
possibilities. Other $L$ and $B-L$ violating processes such as
neutrinoless beta decay $(0\nu\beta\beta)$ - can however
discriminate between Majorana and Dirac neutrinos. The
corresponding effective operator for $\onbb$ is at least dim-9.
Therefore  the scale of new physics can be around the collider
search scale without any serious fine tuning in the parameters in
some models and hence can be probed directly\cite{0nu2betaLR}. $B$
and $B-L$ violating process like $n-\nbar$ oscillation
\cite{Kuzmin:1970nx,MohapatraBminusL,shrock} emerge  from another
dim-9 effective operator. However for high scale $B-L$ violation,
as is the rule for SUSY GUTs, such operators are highly suppressed.

The NMSGUT is a SUSY GUT model based on SO(10) gauge symmetry with
all the fermions (including right handed neutrino) of one
generation  residing in a single 16-plet spinor supermultiplet of
Spin(10). Higgs fields which break gauge symmetry and give masses
reside in chiral supermultiplets that are $\bm{10}$, $\bm{120}$,
$\bm{\overline{126}}$ $(+\bm{126})$, $\bm{210}$  representations
of SO(10). Due to the strong constraints on possible renormalizable
SO(10) cubic invariants this model enjoys a minimal number of
parameters versus all competing viable models.

The complete NMSGUT superpotential is \cite{aulmoh, ag1, ag2, nmsgut},
  {\small \bea {\cal W}&=&\left(h_{AB}
\bm{10}+\frac{f_{AB}}{5!} \bm{\overline{126}} +\frac{g_{AB}}{3!}
\bm{120}\right)\bm{16}_A \bm{16}_B \nn &+&\frac{M_H}{2} \bm{10}^2+\frac{m_\Theta}{2
(3!)} \bm{120}^2 +\frac{M_{\Sigma}}{5!} \bm{126.\overline{126}}\nn
&+&\frac{M_\Phi}{4!}\bm{210.210} + \frac{\lambda}{4!} \bm{210}^3
+\frac{\eta}{4!} \bm{210.126.\overline{126}}\nn &+&
\frac{1}{2(3!)}\bm{120.210.}(\zeta \bm{126}+\overline{\zeta} \bm{\overline{126}})
+ \frac{\kappa}{3!} \bm{10.120.210} \nn &+& \frac{1}{4!}\bm{10.210.}(\gamma
\bm{126} + \overline{\gamma} \bm{\overline{126}})+\frac{\rho}{4!}
\bm{120.120.210} \label{eq:so10super} \eea } where $A,B=1,2,3$,
$h=h^T,f=f^T$ and $g=-g^T$.   The decomposition of these
representations into SM irreps gives  26 types of  distinct
representations which were therefore conveniently labelled
alphabetically\cite{ag1}. The decomposition of tensorial
representations is detailed in \cite{ag1,ag2,nmsgut} together with
their mass matrices for mixed and unmixed states. We will adopt
the same nomenclature here.

For deriving the effective superpotential it is convenient to
divide fields into three categories   according to their coupling
patterns and  masses after SSB : \green fields are  light with
masses of order $M_Z$ or less and populate the effective theory.
\red fields have large masses $>>M_Z$ but  couple with fields in
the matter 16-plets. \blue fields  do not couple with 16-plets
 and are also superheavy. The pair of MSSM Higgs doublets which
is light is thus  \green and rest (five) of the Higgs type doublet
pairs are \red. Similarly the heavy right handed neutrino is \red.
The  Higgs ($\mathbf{10,120,\oot}$) coupled to matter fields are
called fermion mass (FM) Higgs  and the  rest ($\bm{210}$)  are Adjoint
mode (AM) Higgs. The SM decomposition of ${\cal W}_{FM}$ is given
in eq.(58,60) of \cite{ag2} and eq.(5.5) of \cite{nmsgut}. The
superpotential  has \green-\green-\green (GGG) terms containing SM
fermions with either of the light Higgs pair, \green-\green-\red (GGR)
containing two SM fermions and one heavy FM higgs field   or SM
fermion-light Higgs-RH neutrino. ${\cal W}_{AM}$ contributes a
large number of cubic {\blue} fields interacting with {\blue} or {\red}
fields as well as BB and RR masses, but also contains cubic
interactions with {\green}  (light Higgs fields). However the light
Higgs (by definition) lack superpotential mass terms.
\begin{table}
\begin{center}
\begin{tabular}{|p{1.5cm} |p{4.3cm}|}
\hline
\green & $\gubar$, $\gdbar$, $\gQ$, $\gebar$, $\gL$, $\gH$, $\gHbar$ \\
\hline
\red & $\rnubar$,\,$\rA, \rC,\rD, \rE, \rF, \rG, \rh, \rJ, \rK$, $\rL, \rM, \rN, \rO, \rP, \rt, \rW$ \\
\hline
\blue & $\bB, \bI, \bQ, \bR, \bS, \bU, \bV, \bX, \bY, \bZ$ \\
\hline
\end{tabular}
\caption{Nature of the fields} \label{tab:fieldshade}
\end{center}
\end{table}
\begin{eqnarray}
{\cal W}^{GGG}_{FM}&=& \gHbar \left[
-\frac{4}{\sqrt{2}}h_{AB}U^{h\dagger}_{11}(\gdbar_A \gQ_{B}
 + \gebar_A \gL_B ) \right. \nn &+&
4{\sqrt{2}}\frac{i}{\sqrt{3}} f_{AB}  U^{h\dagger}_{12}
(\gdbar_{A}\gQ_{B} - 3 \gebar_{A} \gL_{B}) \nn &+&
2\sqrt{2}g_{AB}U^{h\dagger}_{15} (\gdbar_A \gQ_B + \gebar_A \gL_B)  \nn
&-&2\left.\sqrt{2}g_{AB}\frac{i}{\sqrt{3}}U^{h\dagger}_{16} ( \gdbar_A \gQ_B
- 3\gebar_A \gL_B ) \right] \nn &+& \gH \big (\gubar_A
\gQ_B \big) \left[2\sqrt{2}h_{AB}V^h_{11} - 4{\sqrt{2}}
{i\over\sqrt{3}}f_{AB}  V^h_{21} \right. \nn &-&\left.
2\sqrt{2}g_{AB}V^h_{51} +2\sqrt{2}g_{AB}\frac{i}{\sqrt{3}}V^h_{61}
\right] \label{eq:gggfm}
\end{eqnarray}
The matrices $U^{\Phi},V^{\Phi}$ diagonalize Higgs field
superpotential masses   as $ U^\dagger M V= M_{diag}$ with the
alphabetic superscripts indicating which of the 13 mixing types is
involved (see \cite{nmsgut,bstabhedge} for details).  ${\cal
W}^{GGR}_{FM}$ has two light matter fields  and one heavy Higgs
(\red) field or else  light matter, light Higgs, and  RH neutrino:
{\small
\begin{eqnarray}
{\cal W}^{GGR}_{FM}&=&\Gamma^{\tbar\ubar\dbar}_{\hat{j}AB} \rtbar_{\hat{j}}
\epsilon\gubar_A \gdbar_B
+\Gamma^{\tbar QL}_{\hat{j}AB}\rtbar_{\hat{j}} \gQ_A \gL_B +
\Gamma^{tQQ}_{\hat{j}AB} \rt_{\hat{j}} \frac{\epsilon}{2}\gQ_A \gQ_B  \nn
&+&\Gamma^{t\ubar\ebar}_{\hat{j}AB}{\rt}_{\hat{j}}\gubar_A \gebar_B
+\Gamma^{\Cbar\dbar Q}_{\hat{j}AB}{\rCbar}_{\hat{j}}\gdbar_A \gQ_B
+\Gamma^{C\ubar Q}_{\hat{j}AB}{\rC}_{\hat{j}}\gubar_A \gQ_B \nn &+&
\Gamma^{D\ubar L}_{\hat{j}AB}{\rD}_{\hat{j}}\gubar_A \gL_B
+\Gamma^{\Dbar \ebar Q}_{\hat{j}AB}{\rDbar}_{\hat{j}}\gebar_A \gQ_B +
\Gamma^{E\dbar L}_{{\hat{j}}AB}{\rE}_{\hat{j}}\gdbar_A \gL_B \nn &+&
\Gamma^{\Abar\ebar\ebar}_{AB}{\rAbar}\gebar_A \gebar_B +
\Gamma^{\Wbar QQ}_{AB}{\rWbar} \gQ_A \gQ_B
+\Gamma^{\Pbar QL}_{{\hat{j}}AB}{\rPbar}_{\hat{j}} \gQ_A \gL_B \nn
&+&\Gamma^{PQQ}_{{\hat{j}}AB}\frac{\epsilon}{2}{\rP}_{\hat{j}}\gQ_A \gQ_B
+\Gamma^{\Obar LL}_{AB}{\rObar} \gL_A\gL_B
+\Gamma^{K\dbar\ebar}_{{\hat{j}}AB}{\rK}_{\hat{j}}\gdbar_A\gebar_B \nn
&+&\Gamma^{\Kbar\ubar\ubar}_{{\hat{j}}AB}\epsilon
{\rKbar}_{\hat{j}}\gubar_A \gubar_B +\Gamma^{L\ubar\dbar}_{{\hat{j}}AB}
\rL_{\hat{j}}\gubar_A\gdbar_B
+\Gamma^{\Lbar QQ}_{{\hat{j}}AB}{\rLbar}_{\hat{j}}\gQ_A \gQ_B \nn
&+&\Gamma^{FLL}_{{\hat{j}}AB}{\rF}_{\hat{j}}\gL_A\gL_B
+\Gamma^{\Jbar \dbar \dbar}_{{\hat{j}}AB}\epsilon
{\rJbar}_{\hat{j}}\gdbar_A\gdbar_B
+\Gamma^{\hbaar\dbar Q}_{\hat{\bar{k}}AB}{\rhbar}_{\hat{\bar{k}}}\gdbar_A \gQ_B
\nn &+&\Gamma^{\hbaar\ebar L}_{\hat{\bar{k}}AB}
{\rhbar}_{\hat{\bar{k}}}\gebar_A \gL_B
+\Gamma^{h\ubar Q}_{\hat{\bar{k}}AB}{\rh}_{\hat{\bar{k}}}\gubar_A \gQ_B
+\Gamma^{h{\nubar}L}_{AB}{\gH}\rnubar_A \gL_B \nn &+&
\Gamma^{N\dbar \dbar}_{AB}\rN\gdbar_A \gdbar_B+
\Gamma^{M\ubar\ubar}_{AB}\rM\gubar_A \gubar_B \label{eq:rggfm}
\end{eqnarray}
} The coefficients $\Gamma^{ABC}_{\hat{a}bc}$ are expressed in
terms of GUT parameters by equations like
\begin{eqnarray}
\Gamma^{\tbar\ubar\dbar}_{\hat j AB}&=&\left[2\sqrt{2}h_{AB}U^{t\dagger}_{\hat j 1}
-4\sqrt{2}f_{AB}U^{t\dagger}_{\hat j 2}+4i\,g_{AB}U^{t\dagger}_{\hat j 7}\right] \nnu
\Gamma^{M\ubar\ubar}_{AB}&=& 4\sqrt{2}f_{AB} \label{eq:gammadef}
\end{eqnarray}
and complete expressions will be reported in \cite{aulawasrikri}. The indices
$A,B$ run over three flavor generations of fermions. The hat over the indices of
the \red fields is to indicate that they are in mass diagonal basis. The index
over heavy Higgs doublets $\rh~ (\rhbar)$ starts from 2 and is distinguished by
a bar over it.

Similarly ${\cal W}^{GRR}_{FM}$ has one \green and two \red
fields out of which one is heavy neutrino.
\begin{eqnarray}
{\cal W}^{GRR}_{FM}&=&\Lambda^{t\dbar\nubar}_{{\hat{j}}AB}
\rt_{\hat{j}} \rnubar_B  \gdbar_A
+\Lambda^{\Ebar{\nubar}Q}_{{\hat{j}}AB}{\rEbar}_{\hat{j}}\rnubar_A \gQ_B
+\Lambda^{\Fbar\ebar\nubar}_{{\hat{j}}AB}\rFbar_{\hat{j}}\gebar_A\rnubar_B
\nn &+&\Lambda^{J\ubar\nubar}_{{\hat{j}}AB}
\rJ_{\hat{j}}\gubar_A\rnubar_B
+\Lambda^{h{\nubar}L}_{\hat{\bar{k}}AB}{\rh}_{\hat{\bar{k}}}\rnubar_A \gL_B
\label{eq:rrgfm}
\end{eqnarray}
The coefficients $\Lambda^{ABC}_{\hat{a}bc}$ are expressed
\cite{aulawasrikri} in terms of GUT parameters by equations like
\begin{eqnarray}
\Lambda^{t\dbar\nubar}_{\hat j AB}&=&\left[-2\sqrt{2}h_{AB}V^t_{1\hat j}
+4\sqrt{2}f_{AB}V^t_{2\hat j}
-8 i f_{AB}V^t_{4\hat j}\right.\nn
&&\left. + 4 g_{AB}V^t_{6\hat j}+4 i g_{AB}V^t_{7\hat j}\right]\label{eq:lambdadef}
\end{eqnarray}
  ${\cal W}^{RRR}_{FM}$ has  three \red fields and therefore
  requires two RH neutrinos.
\begin{eqnarray}
{\cal W}^{RRR}_{FM}=-8if_{AB} \rG_5\rnubar_A\rnubar_B  \label{eq:rrrfm}
\end{eqnarray}
Similarly in the adjoint mode we have
\begin{eqnarray}
{\cal W}^{GGR}_{AM}&=&\Upsilon^{O} \gH\gH\rO+\Upsilon^{\bar{O}} \gHbar\gHbar\rObar
+ \Upsilon^{S} \gH\gHbar \rS \nn
&+&\Upsilon^{G}_{\hat{j}} \gH \gHbar \rG_{\hat{j}}
+\Upsilon^{\bar{F}}_{\hat{j}} \gH\gH \rFbar_{\hat{j}}
+\Upsilon^{F}_{\hat{j}} \gHbar\gHbar \rF_{\hat{j}}\label{eq:ggram}
\end{eqnarray}
Note that, as first shown in \cite{ag2} the $\gH\gH\rO$,
$\gHbar\gHbar\rObar$ novel terms are the basis of the Type II
seesaw in the MSGUT due to the $RGG$ term $\rObar\gL\gL$ in eq.(\ref{eq:rggfm}).
The $M_O\rObar\rO$ mass term implies $\left<\rObar\right>\sim \left<\gH\gH\right>/M_O$
leading to Type~II masses for neutrinos. The coefficients $\Upsilon^{A}_{a}$ are
expressed \cite{aulawasrikri} in terms of GUT parameters by equations like
\begin{eqnarray}
\Upsilon^{O}&=& \eta 2\sqrt{3}V^h_{2\hat 1}V^h_{4\hat 1}
+i\zeta\sqrt{3}V^h_{6\hat 1}V^h_{4\hat 1}
+\zeta V^h_{5\hat 1}V^h_{4\hat 1}\label{eq:upsilono}
\end{eqnarray}

Similarly
\begin{eqnarray}
{\cal W}^{GRR}_{AM}&=&\Omega^{h\Fbar}_{\hat{\bar{k}}\hat{j}}\gH
\rh_{\hat{\bar{k}}} \rFbar_{\hat{j}}
+\Omega^{\hbaar F}_{\hat{\bar{k}}\hat{j}} \gHbar
\rhbar_{\hat{\bar{k}}} \rF_{\hat{j}}
+\Omega^{E\Jbar}_{\hat{i}\hat{j}}\gH \rE_{\hat{i}}\rJbar_{\hat{j}} \nn
&+& \Omega^{\Ebar J}_{\hat{i}\hat{j}}\gHbar \rEbar_{\hat{i}}\rJ_{\hat{j}}
+\Omega^{t\Ebar}_{\hat{i}\hat{j}}\gH \rt_{\hat{i}}\rEbar_{\hat{j}}
+{\Omega}^{\tbar E}_{\hat{i}\hat{j}}\gHbar \rtbar_{\hat{i}}\rE_{\hat{j}} \nn
&+& \Omega^{\Ebar P}_{\hat{i}\hat{j}}\gH \rP_{\hat{i}}\rEbar_{\hat{j}}
+{\Omega}^{E\Pbar}_{\hat{i}\hat{j}}\gHbar\rPbar_{\hat{i}}\rE_{\hat{j}}
+\Omega^{\hbaar G}_{\hat{\bar{k}}\hat{j}}\gH \rhbar_{\hat{\bar{k}}}\rG_{\hat{j}} \nn
&+&\Omega^{hG}_{\hat{\bar{k}}\hat{j}} \rh_{\hat{\bar{k}}} \gHbar \rG_{\hat{j}}
+\Omega^{J\Dbar}_{\hat{i}\hat{j}} \gH \rJ_{\hat{i}}\rDbar_{\hat{j}}
+{\Omega}^{\Jbar D}_{\hat{i}\hat{j}}\gHbar \rJbar_{\hat{i}}\rD_{\hat{j}} \nn
&+& \Omega^{\hbaar S}_{\hat{\bar{k}}} \gH \rhbar_{\hat{\bar{k}}} \rS
+{\Omega}^{hS}_{\hat{\bar{k}}}\rh_{\hat{\bar{k}}}\gHbar \rS
+\Omega^{hO}_{\hat{\bar{k}}} \gH \rh_{\hat{\bar{k}}} \rO \nn
&+&{\Omega}^{\hbaar \Obar}_{\hat{\bar{k}}} \gHbar \rhbar_{\hat{\bar{k}}}\rObar \label{eq:grram}
\end{eqnarray}

The coefficients $\Omega^{AB}_{\hat{a}\hat{b}}$ are expressed
\cite{aulawasrikri} in terms of GUT parameters by equations like
\begin{eqnarray}
\Omega^{h\Fbar}_{\hat{\bar{j}}\hat k}&=&
\frac{4\eta}{\sqrt{3}}V^{h}_{2\hat 1}V^{h}_{3\hat{\bar{j}}}U^{F\dagger}_{\hat k 2}
+i\eta 2\sqrt{3}V^{h}_{3\hat 1}V^{h}_{4\hat{\bar{j}}}U^{F\dagger}_{\hat k 1} \nn
&-&\frac{i\rho}{3}V^{h}_{5\hat 1}V^{h}_{6\hat{\bar{j}}}U^{F\dagger}_{\hat k 2}
+\frac{i\rho}{\sqrt{3}}V^{h}_{6\hat 1}V^{h}_{4\hat{\bar{j}}}U^{F\dagger}_{\hat k 4} \nn
&+&\frac{2i\zeta}{\sqrt{3}}V^{h}_{6\hat 1}V^{h}_{3\hat{\bar{j}}}U^{F\dagger}_{\hat k 2}
+\zeta V^{h}_{5\hat 1}V^{h}_{3\hat{\bar{j}}}U^{F\dagger}_{\hat k 2} \nn
&+&\zeta\sqrt{3}V^{h}_{3\hat 1}V^{h}_{4\hat{\bar{j}}}U^{F\dagger}_{\hat k 4}
+\bar{\zeta}V^{h}_{5\hat 1}V^{h}_{2\hat{\bar{j}}}U^{F\dagger}_{\hat k 2} \nn
&+&\frac{2i\bar{\zeta}}{\sqrt{3}}V^{h}_{6\hat 1}V^{h}_{2\hat{\bar{j}}}U^{F\dagger}_{\hat k 2}
-\bar{\zeta}\sqrt{\frac{3}{2}}V^{h}_{6\hat 1}V^{h}_{4\hat{\bar{j}}}U^{F\dagger}_{\hat k 1} \nn
&+&\bar{\zeta}\sqrt{3}V^{h}_{2\hat 1}V^{h}_{4\hat{\bar{j}}}U^{F\dagger}_{\hat k 4}
- i\bar{\zeta}V^{h}_{5\hat 1}V^{h}_{4\hat{\bar{j}}}U^{F\dagger}_{\hat k 1}\nn
&+&\left(V^h_{m\hat{1}}V^h_{n\hat{\bar{j}}}
\leftrightarrow V^h_{n\hat{1}}V^h_{m\hat{\bar{j}}}\right)\label{eq:omegahf}
\end{eqnarray}
Similarly we will have ${\cal W}^{RRR}_{AM}$, ${\cal W}^{GRB}_{AM}$ and ${\cal
W}^{RRB}_{AM}$. In AM contributions $G$
can only be a  light Higgs doublet. Going beyond ${\cal W}^{GRB}_{AM}$ is not
required for obtaining the sextic effective potential, and the
relevant operators would be even more hopelessly suppressed. Mass
terms of heavy fields are not given  here because  they acquire
masses both from mass terms and the  superheavy VEVs  of the Higgs
fields and the mass matrices for all such heavy fields are given
in \cite{ag2, nmsgut}. Now using the equations of motion we sequentially
integrate out \blue fields in terms of \red-\green fields, \red
fields in terms of \green-\green fields. At the end of this section
we have shown that the superpotential terms with \blue fields give raise
to sextic or higher dimensional effective operators. The quartic and
quintic effective operator purely emerge from the superpotential terms with one
or two \red fields. Plugging them back in to the superpotential terms
leads to effective \green operators. For
example the low energy equations  of motion for \red fields like RH
neutrino (see eq.(\ref{eq:gggfm})) and   $F$ (see eq.(\ref{eq:rrgfm}) and (\ref{eq:ggram})) give : 
\bea
\rnubar_A&=&-{M^{\bar{\nu}}_{AB}}^{-1}\left[\Gamma^{h{\nubar}L}_{BC} \gH \gL_C
\right] \label{eq:eomnu}\\
\rF_{\hat{j}}&=&-{\cal
F}_{\hat{j}\hat{j}}\left[\Upsilon^{\bar{F}}_{\hat{j}} \gH \gH +
\Lambda^{\Fbar\ebar\nubar}_{\hat{j}AB}\gebar_A \rnubar_B \right]
\label{eq:eomf} \eea where ${\cal F}$ is the   inverse of the
diagonalized $4 \times 4$ mass matrix  for $ \bar F-F$ type
fields, and other coefficients are like \bea \Gamma^{h\nubar
L}_{AB}&=&2\sqrt{2}
h_{AB}V^h_{1\hat{1}}+4i\sqrt{6}f_{AB}V^h_{2\hat{1}} \nn
&-&2\sqrt{2}g_{AB}V^h_{5\hat{1}}-2i\sqrt{6}g_{AB}V^h_{6\hat{1}}\,.
\label{eq:coupeomnu} \eea Although we have called both $\rnubar$
and $\rA,\rC \cdots$ \red (see Table\,\ref{tab:fieldshade}) the
$\rnubar$ masses lie in $10^7-10^{13}$ GeV and are thus much
lighter than typical GUT scales. This difference in scales is set
by the requirement that the Type I seesaw masses match the
observed neutrino masses. Since in the MSGUT we break SO(10)  to
the SM gauge group in a single step, and integrating out right
handed neutrinos in stages increases the complexity without
drastic  effects on the neutrino masses we integrate out all the
right handed neutrinos at the GUT breaking scale as well. An
improved treatment\cite{lindner} would first integrate out only
GUT scale mass   particles calculate renormalization group flow corrected couplings
at the heaviest neutrino scale, integrate out the heaviest
neutrino and repeat till the MSSM corrected by the $d=5$ Weinberg
operator for Majorana neutrino masses was reached. While this
procedure may be followed in numerical work, our intention here is
to indicate  the effects of GUT scale thresholds. We shall show
below that the same  GUT threshold corrections which suppress
proton decay facilitate  the separation of right handed neutrino
masses from GUT scales and hence more natural. Terms coming from
$\rnubar$ substitution are more important than generic \red Higgs
fields.
%Note in eq.(\ref{eq:eomf}) that the Red field $\rF$ contains a term with lighter red field $\rnubar$.
Substituting $\rnubar$ from eq.(\ref{eq:eomnu}) in eq.(\ref{eq:eomf})
we see that term with coefficient $\Lambda^{\Fbar\ebar\nubar}_{\hat{j} AB}$ is
subleading compared to the term
with coefficient $\Upsilon^{\bar{F}}_{\hat{j}}$, and will contribute to quintic operator.
%due to suppression by $(M^{\bar{\nu}})^{-1}$.
Similarly, we can see from eq.(\ref{eq:rrgfm}), the
fields $\rE,\rtbar,\rJbar$ and $\rhbar$ will contain such $\rnubar$ dependent terms.

Now, plugging back these results in ${\cal W}^{GGR}_{FM}$ (eq.(\ref{eq:rggfm})) and
${\cal W}^{GGR}_{AM}$ (eq.(\ref{eq:ggram})) we get the following quartic superpotential
at the leading order {\small
\begin{eqnarray}
&&{\cal W}^4={\cal W}^4_{(1,1)}+{\cal W}^4_{(0,0)}+{\cal W}^4_{(-1,-1)}
+{\cal W}^4_{(0,2)}\nonumber\\
&&{\cal W}^4_{(1,1)}=L_{ABCD}\frac{\epsilon}{2}\gQ_A\gQ_B\gQ_C\gL_D \nonumber \\
&&{\cal W}^4_{(-1,-1)}=R_{ABCD}\epsilon \gebar_A\gubar_B\gubar_C\gdbar_D \nonumber \\
&&{\cal W}^4_{(0,2)}=P^{I}_{AB} (\gL_A\gL_B)(\gH\gH)+P^{II}_{AB}(\gL_A\gH)(\gL_B\gH)
\nonumber\\
&&{\cal W}^4_{(0,0)}=Q^{I}_{ABCD} (\gQ_A\gL_B)\gubar_C\gebar_D + Q^{II}_{ABCD}
(\gQ_A\gdbar_D)(\gQ_B\gubar_C)\nonumber\\
&&\hspace{1cm} +Q^{III}(\gHbar \gH)(\gHbar \gH)+
Q^{IV}(\gHbar \gHbar)(\gH\gH) \label{eq:eff4}
\end{eqnarray}
}\hspace{-.12cm}
where subscripts on the superpotential components denote the $(B, L)$ numbers.
The coefficients $L_{ABCD}$ and $R_{ABCD}$ are given in eq.(5.7 $\&$ 5.8)
of \cite{nmsgut}  and other coefficients are like
\begin{eqnarray}
P^I_{AB}=-\Gamma^{FLL}_{\hat{j}AB}{\cal F}_{\hat{j}\hat{j}}
\Upsilon^F_{\hat{j}}, \label{eq:eff4pi}
\end{eqnarray}
where  $\Gamma^{FLL}_{\hat{j}AB} =
-(2\sqrt{2}g_{AB}) V^F_{4\hat j}$ %\label{eq:gammafll}
and {\small
\begin{eqnarray}
\Upsilon^F_{\hat j}&=&\left[
-\frac{4\eta}{\sqrt{3}}U^{h\dagger}_{\hat 1 2}U^{h\dagger}_{\hat 1 3}V^F_{2\hat j}
-i\eta 2\sqrt{3}U^{h\dagger}_{\hat 1 2}U^{h\dagger}_{\hat 1 4}V^F_{1\hat j}\right. \nn
&+&\left. \frac{i\rho}{3}U^{h\dagger}_{\hat 1 5}U^{h\dagger}_{\hat 1 6}V^F_{2\hat j}
-\frac{i\rho}{\sqrt{3}}U^{h\dagger}_{\hat 1 6}U^{h\dagger}_{\hat 1 4}V^F_{4\hat j}
-\frac{2i\zeta}{\sqrt{3}}U^{h\dagger}_{\hat 1 6}U^{h\dagger}_{\hat 1 3}V^F_{2\hat j}
\right.\nn
&+& \zeta\sqrt{3}U^{h\dagger}_{\hat 1 3}U^{h\dagger}_{\hat 1 4}V^F_{4\hat j}
+i\zeta U^{h\dagger}_{\hat 1 5}U^{h\dagger}_{\hat 1 4}V^F_{1\hat j}
-\zeta U^{h\dagger}_{\hat 1 5}U^{h\dagger}_{\hat 1 3}V^F_{2\hat j} \nn
&-& \zeta\sqrt{3}U^{h\dagger}_{\hat  1 6}U^{h\dagger}_{\hat 1 4}V^F_{1\hat j}
-\frac{2i\bar{\zeta}}{\sqrt{3}}U^{h\dagger}_{\hat 1 6}U^{h\dagger}_{\hat 1 2}V^F_{2\hat j}
-\bar{\zeta}U^{h\dagger}_{\hat 1 2}U^{h\dagger}_{\hat 1 5}V^F_{2\hat j} \nn
&+& \left. \bar{\zeta}\sqrt{3}U^{h\dagger}_{\hat 1 2}U^{h\dagger}_{\hat 1 4}V^F_{4\hat j}
+\bar{\zeta}\sqrt{3}U^{h\dagger}_{\hat 1 2}U^{h\dagger}_{\hat 1 4}V^F_{4j}\right]
\label{eq:unsilonf}
\end{eqnarray}
} and similarly we can find $P^{II}_{AB}$ and
$Q^{I,II,III,IV}_{ABCD}$\cite{aulawasrikri}. As explained above, in the dim-4
superpotential ${\cal W}^{(4)}_{(0,2)}$ there exist operators
which are Type I and II seesaw terms and violate $B-L$. The operator in
${\cal W}^{GGR}_{FM}+{\cal W}^{GGR}_{AM}$ with fields $\rF$, $\rE$,
$\rtbar$, $\rJbar$, $\rhbar$ will also give
rise to contributions to quintic operators since $\rF$, $\rE$,
$\rtbar$, $\rJbar$, $\rhbar$ are ${\cal O}(G^3)$ through
$\rnubar$.

The other quintic effective superpotential terms arise from integrating
out two \red fields from ${\cal W}^{GRR}_{FM}+{\cal W}^{GRR}_{AM}$. There
are three types of quintic operators:
{\small
\begin{eqnarray}
{\cal W}^5&=& {\cal W}^5_{(-1,1)} +{\cal W}^5_{(0,2)} +{\cal W}^5_{(0,0)} \nonumber\\
{\cal W}^5_{(-1,1)}&=& F^{I}_{ABCD}(\gH\gL_A)\epsilon \gdbar_B \gdbar_C \gubar_D
+F^{II}_{ABCD}(\gHbar \gL_A) \epsilon \gdbar_B \gdbar_C\gdbar_D \nonumber \\ %\\
W^5_{(0,2)}&=& G^{I}_{ABCD}\gdbar_A(\gH\gQ_B)(\gL_C\gL_D)
+G^{II}_{ABCD}\gdbar_A (\gH\gL_B)(\gQ_C\gL_D) \nn
&& +G^{III}_{ABCD}\gebar_A(\gH\gL_B)(\gL_C\gL_D)\nonumber \\
 W^5_{(0,0)}&=& H^{I}_{AB}\gebar_A(\gH\gL_B)(\gHbar\gHbar)
+H^{II}_{AB}\gebar_A(\gHbar \gL_B)(\gH\gHbar)\nonumber\\
&&+H^{III}_{AB}\gubar_A(\gHbar\gQ_B)(\gH\gH)
+H^{IV}_{AB}\gubar_A(\gH\gQ_B)(\gH\gHbar) \nonumber \\
&&+H^{V}_{AB}\gdbar_A(\gHbar\gQ_B)(\gH\gHbar)
+H^{VI}_{AB}\gdbar_A(\gH\gQ_B)(\gHbar\gHbar)\label{eq:eff5}
\end{eqnarray}
and the coefficients are like
\begin{eqnarray}
%&&F^{I} \sim \frac{ (f/g/h)^3}{(M_t/M_J)M_\nu} + \cdots \nn
%&&F^{II} \sim \frac{ (f/g/h)^2(\eta/\zeta/\rho...)}{(M_EM_J)} + \cdots \nonumber
F^I_{ABCD} &=&\left[\Gamma^{\tbar\ubar\dbar}_{\hat{j}DC}{\cal T}_{\hat{j}\hat{j}}
\Lambda^{t\dbar\nubar}_{\hat{j}BE} + \Gamma^{\Jbar\dbar\dbar}_{\hat{j}BC}
{\cal J}_{\hat{j}\hat{j}} \Lambda^{J\ubar\nubar}_{\hat{j}DE}\right] ({\cal M}_{\bar{\nu}}^{-1})_{EF}
\Gamma^{h{\nubar}L}_{FA} \nn
&-& \left[\Omega^{J\Dbar}_{\hat{i}\hat{j}}{\cal J}_{\hat{i}\hat{i}}
\Gamma^{\Jbar\dbar\dbar}_{\hat{i}BC}{\cal D}_{\hat{j}\hat{j}}
\Gamma^{D\ubar L}_{\hat{j}DA}\right] -
\left[ \Omega^{t\Ebar}_{\hat{i}\hat{j}}{\cal T}_{\hat{i}\hat{i}}
\Gamma^{\tbar\ubar\dbar}_{\hat{i}DB}{\cal E}_{\hat{j}\hat{j}}\Gamma^{E\dbar L}_{\hat{j}CA}
\right] \nn
F^{II}_{ABCD}&=&\left[-{\Omega}^{\Ebar J}_{\hat{i}\hat{j}}{\cal E}_{\hat{i}\hat{i}}
\Gamma^{E\dbar L}_{\hat{i}BA}{\cal J}_{\hat{j}\hat{j}}\Gamma^{\Jbar\dbar\dbar}_{\hat{j}CD}
\right] \label{eq:eff5coeff}
\end{eqnarray}
}\noindent where $\cal T$, $\cal J$, $\cal D$ and $\cal E$ are the
inverse of mass matrices $M_t$, $M_J$, $M_D$ and $M_E$
respectively. Since $M_{\bar{\nu}}<< M_{t,J,D,E}$, the first term of coefficient
$F^{I}_{ABCD}$ in eq.(\ref{eq:eff5coeff}) is most significant. Patterns of the
coefficients $\Gamma,\Lambda$ and $\Omega$ are given in
eq.(\ref{eq:gammadef},~\ref{eq:coupeomnu}), (\ref{eq:lambdadef}) and (\ref{eq:omegahf}).

Most interestingly, here we have $B, L$ and $B-L$ violating
operators ${\cal W}^5_{(-1,1)}$. They open $B-L$ violating channels to nucleon decay.
The operator $(\gH\gL_A)\epsilon \gdbar_B \gdbar_C \gubar_D$ has a contribution
from ${\cal W}^{GRR}_{FM}$ with one \red field being the right handed
neutrino, and was discussed in \cite{babumohap, nathraza}. It also gets a
contribution from ${\cal W}^{GRR}_{AM}$
through $\gH\rJ\rDbar$ and $\gH\rt\rEbar$ terms (see eq.(\ref{eq:ggram})).
In addition to that we have a new quintic operator
$(\gHbar \gL_A) \epsilon \gdbar_B \gdbar_C\gdbar_D$ - with coefficient
$F^{II}_{ABCD}$ as given in eq.(\ref{eq:eff5coeff}) - which arises from
$\gHbar\rEbar\rJ$ of ${\cal W}^{GRR}_{AM}$. We will emphasize its importance in
Section\,\ref{sec:effbl}.

At the next to leading order ${\cal W}^{GRR}_{FM}+{\cal W}^{GRR}_{AM}$
also contributes to sextic operator.
From the discussion above we can easily infer that the leading order
effective superpotential emerging from ${\cal W}^{RRR}$ will be sextic
in order. Similarly, integrating out the \blue field effectively gives
$B\sim GR/M_B+RR/M_B$ which, on replacing the \red field, becomes
\bea
B\sim \frac{GGG}{M_R M_B}+\frac{GGGG}{M_R^2M_B} + {\cal O}(G^5/M_X^4).
\eea
Therefore, we see that the leading effective operators emerging from
${\cal W}^{GRB}_{AM}$ are also sextic. The effective operator emerging
from ${\cal W}^{RRB}_{AM}$ are septic or higher. We have listed all
the sextic operators emerging from ${\cal W}^{GRR}_{FM}$, ${\cal W}^{GRR}_{AM}$,
${\cal W}^{RRR}_{AM}$ and ${\cal W}^{GRB}_{FM}$ in Table~\ref{tab:dim7},
and have separated out the class of sextic operators emerging only from
$W^{GRB}_{FM}$ in Table~\ref{tab:dim7a}.

\section{Threshold enhancements and proton decay}\label{sec:threnh}
The problem of fast  proton decay in NMSGUT due to quartic terms
in the effective superpotential, which has accompanied
Supersymmetric GUTs from the beginning\cite{dim5decay} can be
rectified if GUT scale threshold corrections to matter fermion and
MSSM Higss vertices are incorporated\cite{bstabhedge}. Due to the
large number  of heavy chiral multiplets a large wave function
renormalization arises driving the light Higgs fields close to
dissolution ($Z_{H,\bar H}\simeq 0$), which modify MSSM-GUT Yukawa
matching condition.   Rewriting the renormalized Kinetic term
\begin{eqnarray}
%{\cal L}_{Yuk}&=&\bar{f}^TY_ffH_f + h.c. \nn
{\cal L}_{Kin}&=&\bigg[ \sum_{A,B}\bar{f}_A^\dagger (Z_{\bar{f}}
)_A^B\bar{f}_B+{f}_A^\dagger
(Z_{f})_A^Bf_B   \nonumber\\
&& + H^\dagger Z_H H + \bar{H}^\dagger Z_{\bar{H}} \bar{H}\bigg]_D + \cdots
\label{eq:kinlag}
\end{eqnarray}
with  canonical normalization  requires the
transformation\cite{bstabhedge}
\begin{eqnarray}
f=U_{Z_f}\Lambda_{Z_f}^{-1/2} \widetilde{f}, &&
\bar{f}=U_{Z_{\bar{f}}}\Lambda_{Z_{\bar{f}}}^{-1/2}\widetilde{\bar{f}}, \nonumber\\
H=\widetilde{H}/\sqrt{Z_H}, &&
\bar{H}=\widetilde{\bar{H}}/\sqrt{Z_{\bar{H}}}\,, \label{eq:fieldthr}
\end{eqnarray}
where $\Lambda_Z=U^\dagger Z U$ is diagonal, and the Yukawa couplings become
\begin{equation}
\hspace{2cm} \widetilde{Y}_f=\Lambda^{-1/2}_{Z_{\bar{f}}}U^T_{Z_{\bar{f}}}
\frac{Y_f}{\sqrt{Z_{H_f}}}U_{Z_f}\Lambda^{-1/2}_{Z_f}\,, \label{eq:yukawathr}
\end{equation}
and these $\widetilde{Y}_f$ are to be matched with MSSM Yukawa
couplings (not the original tree level $Y_f$). The crucial point is
that  if $Z_H$ is small, the Spin(10) Yukawa couplings
$\{h,f,g\}_{AB}$  required to match the SM Yukawa couplings are
small compared to what they would be in the absence of threshold
corrections. Since   the coefficients of quartic superpotential
baryon decay operators depend on $\{h,f,g\}_{AB}$ and not on
$Z_{H,\bar H}$ it follows that these operators can be suppressed
by one power, and the baryon decay rate by two powers of
$Z_{H,\bar H}$. For small  enough $Z_{H,\bar H}$  this pushes the
proton lifetime into an acceptable range, specially when one
recalls that the freedom to utilize the sfermion diagonalization
matrices can significantly soften\cite{bps} the $B$
violation problem\cite{piercemurayama}.

\section{$B-L$ violating processes}\label{sec:effbl}
Rectification of the quartic operator proton decay  problem is not
the only outcome of threshold corrections. Every effective
operator may get corrected. Let us first look for seesaw operators
which give $B-L$ violating processes.

\noindent
{\bf \underline{Type~I seesaw:}} \\
The relevant part of the superpotential is
\begin{eqnarray}
{\cal W} &\supset & \Gamma^{h \nubar L}\gH\rnubar \gL +\frac{1}{2}M_{\nubar}
\rnubar\rnubar~. %\nn
%&=& \tilde{\Gamma}^{h \nu L}_{AB} \tilde{\gH} {\color{red}\tilde{\rnubar}}_A
%\tilde{\gL}_B+\frac{1}{2}M_{\bar\nu} Z_{\bar\nu}
%{\color{red}\tilde{\rnubar} \tilde{\rnubar}}  \nonumber \\
\end{eqnarray}\label{eq:type1super}
Integrating out the \red field $\rnubar$ produces Type~I seesaw quartic operator
$$
{\cal O}_{\nu} \simeq -\left(\Gamma^{h\nubar L}\right)^2\frac{(\gL\gH)^2}{2 M_{\nubar}}
$$
which can be written in the canonical basis as

\begin{eqnarray}
{\cal O_\nu}\simeq -\left(\tilde{\Gamma}^{h \nubar L}\right)^2
 {\frac{(\tilde{\gL}\tilde{\gH})^2}{2 {\tilde{M}}_{\nubar} }}
\label{eq:type1}
\end{eqnarray}
where $\Gamma^{h \nubar L}_{AB}$ is given in
eq.\,(\ref{eq:coupeomnu}),
$\tilde{\Gamma}=\Gamma/\sqrt{Z_H\Lambda_{Z_L}\Lambda_{Z_{\nubar}}}$
  is the $\sqrt{Z_H}$ boosted tree level Yukawa coupling for the right handed
  neutrinos and $\tilde{M}_{\nubar}=M_{\nubar}/\Lambda_{Z_{\nubar}}$.  Since
typically \cite{bstabhedge} the   wave function renormalization in
$ \tilde{\Gamma}^{h \nubar L}$   only brings the suppressed SO(10)
Yukawa couplings  $h, f, g$ to ordinary MSSM magnitudes  and  the
required size of $\tilde{M}_{\nubar}$  and therefore $f_{AB}$ is
fixed by the observed neutrino masses  the Type~I seesaw acquires
an enhancement  only in the sense that the  \boot  coupling $
f_{AB}$ that enters  $ \tilde{M}_{\nubar}$ is the unboosted one :
making it somewhat easier to achieve righthanded neutrino
 masses much smaller than the GUT scale, while the coupling  $ f_{AB}$   that enters the Yukawa
 couplings that determine the Dirac masses of SM fermions is the boosted one. Thus the rescalings with small $\sqrt{Z_{H}}$
 make  the realistic Type I Seesaw more realistic and explain also the wide
 divergence of the right handed neutrino mass and the GUT scale.\\

\noindent {\bf \underline{Type~II seesaw:}}\\
On the other hand we have\cite{ag2,nmsgut}
\begin{eqnarray}
W &\supset & \Gamma^{\Obar LL}_{AB}\rObar \gL_A \gL_B +
\Upsilon^{\bar{O}} \gHbar \gHbar\rObar+\Upsilon^{O}\gH\gH\rO
+ M_O \rObar \rO \nonumber \\
{\cal O}_\nu & \simeq & -\Gamma^{\Obar LL}_{AB}\Upsilon^O
\frac{(\gL_A\gL_B)(\gH\gH)}{M_O} \nn
&= & -\tilde{\Gamma}^{\Obar LL}_{AB}\Upsilon^O
\frac{(\tilde{\gL}_A\tilde{\gL}_B)(\tilde{\gH}\tilde{\gH})}{\sqrt{Z_H} M_O}
\label{eq:type2}
\end{eqnarray}
where $\Gamma^{\Obar LL}_{AB}=4\sqrt{2}f_{AB}$, $\Upsilon^O$ is given in eq.\,(\ref{eq:upsilono})
and $\tilde{\Gamma}^{\Obar LL}=\Gamma^{\Obar LL}/(\Lambda_{Z_L}\sqrt{Z_H})$.
This operator gets $ Z_H^{-1/2}$ enhancement after counting $Z_H^{-1/2}$ towards
bringing $f_{AB}$ to MSSM Yukawa levels. Therefore, for small $Z_H$, the
Type~II seesaw  has a better chance of  making  a significant contribution to the neutrino
masses. This might have  revived  a possibility that had been dismissed with
some effort\cite{gmblm,blmdm,bert3}, but in practice numerical fits tend to  show that the
  boost still leaves  the Type II seesaw contribution short of, or barely at,  the   milli-eV range. \\

\noindent{\bf \underline{Nucleon decays ($B-L\neq 0$):}}
In Table\,\ref{tab:dim5} the quartic effective operators are
listed. The quartic baryon number violating operators for nucleon
decay preserve  $B-L$. The $B-L$ violating processes like
$p\rightarrow \nu K^+, n\rightarrow e^-K^+, e^-\pi^+$ can only
arise via a different mechanism. However at the quintic operator
level (see the operators listed in Table\,\ref{tab:dim6}),
we have two operators which violate $B, L$ and $B-L$.
Since the VEV of $\gH$ field picks the neutrino in the operator
$\epsilon \gdbar_A \gdbar_B \gubar_C (\gL_D \gH)$, it allows
neutron decay process with neutral leptons $n\rightarrow\nu K^0$ only. But, in
the new operator $\epsilon \gdbar_A \gdbar_B \gdbar_C (\gL_D \gHbar)$
the VEV of $\gHbar$ field picks the charged lepton allowing
$n\rightarrow e^-K^+$. Therefore, this operator seems
novel and needs to be included in the discussion of $B-L$
violating  Baryogenesis supported by SO(10)
GUTs \cite{babumohap}.
The life time for  processes
due to $ {\cal W}^5_{(-1,1)}$ is proportional to
$|F^{I,II}|^{-2}$ (see eq.(\ref{eq:eff5coeff})). Thus we estimate that
the additional presence
of a light Higgs field in the operator implies that the
suppression factor on the rate \emph{relative to the suppressed
threshold corrected  quartic  operator rate}   is of order
$\left<\gH\right>^2/(M_X^2 Z_H)$ so that at best
\begin{eqnarray}
\tau_{n\to \nu K^0,~e^- K^+} &\sim & Z_H 10^{58} \,yrs,
\label{eq:lifebl1}
%&\sim & \Gamma_{n\to \nu K^0} \sim 0.002 (F^I)^2\left(\frac{v_u}{M_S}\right)^2
%\left(\frac{\alpha_S}{\pi}\right)^2
\end{eqnarray}
The effect of threshold corrections relative to  the additional
high scale suppression is thus   modest.  Thus in the NMSGUT the
lifetime for these $B-L$ violating modes  is too large for direct
detection. Its significance for $B-L$ mediated Baryogenesis needs
detailed evaluation in view of the additional possibilities for CP
violation, or exceptional parameter combinations in these
operators.

\section{Conclusions}\label{sec:conclusion}
We have   presented the gist of our results concerning the
effective superpotential  emerging from NMSGUT up to terms sextic
in the light fields. We have also discussed  how  the GUT scale
threshold corrections which rescue\cite{bstabhedge} the
phenomenologically successful
NMSGUT\cite{ckn,aulmoh,abmsv,ag2,nmsgut} impact various processes
including $B-L$ violating processes. We noted that the Type II
seesaw neutrino masses are boosted  sufficiently by threshold
corrections to require inspection for significance in detailed
fits. Examples of expressions for the coefficients of the
effective operators were given here   but detailed results  will
be presented  elsewhere\cite{aulawasrikri} due to their length. As
is  always true in Supersymmetric R-parity preserving GUTs, since
the survival hypothesis fails due to intermediate scales resulting
in large pseudo-goldstone multiplets ruining gauge coupling 
unification \cite{aulmoh}, the $B-L$, $\rm SU(2)_R$
violation scales are required to be quite near the GUT scale.
Hence the novel higher dimension operators, failing the discovery
of anomalously enhanced coefficients for some special case, are
severely suppressed, in spite of the enhancement by the near
dissolution values $(Z_{H,\bar H}<<1)$ of the light Higgs
renormalization factors. A detailed evaluation of the feasibility
of $B-L$ violating Baryogenesis will be considered once a fully
satisfactory  fit taking account of off diagonal coupling matching
at the Susy breaking scale is completed\cite{aulawasrikri}.

\section*{Appendix}

In this Appendix we present tables of the different types of
effective superpotential operators, upto sextic order,  that  we
found by integrating out heavy fields using the
superpotential (momentum independent) equations of motion.

\begin{table}[h!]
\begin{center}
\begin{tabular}{|p{.2cm} |p{2.3cm}|p{.45cm}|p{.45cm}|p{0.8cm}|p{0.8cm}|}
\hline
\multicolumn{6}{|c|}{${\cal W}={\cal O}^{d=4}/M_X$} \\
\hline
&Operators ${\cal O}_i$ & $B$& $L$ & $B-L$ & Enh.\\
\hline
 1&$\epsilon Q_A Q_B Q_C L_D$ & $~~1$ & $~~1$ & $~~0$ & No\\
\hline
2&$Q_A L_B\ubar_C\ebar_D$ & $~~0$   & $~~0$ & $~~0$& No \\
\hline
3&$Q_AQ_B\ubar_C\dbar_D$ & $~~0$   & $~~0$ & $~~0$& No\\
\hline
 4&$\epsilon \ebar_A \ubar_B \ubar_C \dbar_D$ & $-1$ & $-1$ & $~~0$ & No \\
\hline
\multirow{2}{*}{5} & $(L_AL_B)(HH)$,
&\multirow{2}{*}{$0$}&\multirow{2}{*}{$2$}&\multirow{2}{*}{$-2$} & \multirow{2}{*}{$1/{Z_H}$}
\\ & $(L_AH)(L_BH)$ &&&&\\
\hline
\multirow{2}{*}{6} & $(\bar{H}H)(\bar{H}H)$,
&\multirow{2}{*}{$0$}&\multirow{2}{*}{$0$}&\multirow{2}{*}{$0$}
& \multirow{2}{*}{$1/{Z^2_H}$}
\\ & $(HH)(\bar{H}\bar{H})$ &&&&\\
\hline
\end{tabular}
\caption{ Quartic  Effective  Superpotential Operators from
NMSGUT.}\label{tab:dim5}
\end{center}
\end{table}

\begin{table}[t!]
\begin{center}
\begin{tabular}{|p{.3cm} |p{2.55cm}|p{.5cm}|p{.5cm}|p{0.8cm}|p{1.cm}|}
\hline
\multicolumn{6}{|c|}{${\cal W}={\cal O}^{d=5}_i/M^2_X$} \\
\hline
& Operators ${\cal O}_i$ & $B$& $L$ & $B-L$ & Enh. \\
\hline
 1&$\epsilon \dbar_A\dbar_B\ubar_C(L_DH)$ & $-1$ & $~~1$ & $-2$ & $1/\sqrt{Z_H}$\\
\hline
 2&$\epsilon \dbar_A\dbar_B\dbar_C(L_D\bar{H})$ & $-1$ & $~~1$ & $-2$ & $1/\sqrt{Z_H}$ \\
\hline
\multirow{2}{*}{3} & $(Q_AL_B)(L_CH)\dbar_D$,
&\multirow{2}{*}{$0$}&\multirow{2}{*}{$2$}&\multirow{2}{*}{$-2$} & \multirow{2}{*}{$1/\sqrt{Z_H}$}
\\ & $(L_AL_B)(Q_CH)\dbar_D$ &&&&\\
\hline
4&$(L_AL_B)(L_CH)\ebar_D$ & $~~0$   & $~~2$ & $-2$ & $1/\sqrt{Z_H}$ \\
\hline
\multirow{2}{*}{5} & $(L_AH)(\bar{H}\bar{H})\ebar_B$,
&\multirow{2}{*}{$0$}&\multirow{2}{*}{$0$}&\multirow{2}{*}{$0$} & \multirow{2}{*}{$1/\sqrt{Z^3_H}$}
\\ & $(L_A\bar{H})(H\bar{H})\ebar_B$ &&&&\\
\hline
\multirow{2}{*}{6} & $(Q_A\bar{H})(HH)\ubar_B$,
&\multirow{2}{*}{$0$}&\multirow{2}{*}{$0$}&\multirow{2}{*}{$0$} & \multirow{2}{*}{$1/\sqrt{Z^3_H}$}
\\ & $(Q_AH)(H\bar{H})\ubar_B$ &&&&\\
\hline
\multirow{2}{*}{7} & $(Q_A\bar{H})(H\bar{H})\dbar_B,$
&\multirow{2}{*}{$0$}&\multirow{2}{*}{$0$}&\multirow{2}{*}{$0$} & \multirow{2}{*}{$1/\sqrt{Z^3_H}$}
\\ & $(Q_AH)(\bar{H}\bar{H})\dbar_B$ &&&&\\
\hline
\end{tabular} \\
\vspace*{.3cm}
%$W^{d=5}+$\,dressing $\to {\cal L}^{d=6}$.
\end{center}
\caption{Quintic Effective  Superpotential Operators from
NMSGUT.}\label{tab:dim6}
\end{table}

\renewcommand{\arraystretch}{1.5}
\begin{table}[t]
\begin{center}
\begin{tabular}{|p{.4cm} |p{3.8cm}|p{.45cm}|p{.45cm}|p{0.75cm}|}
\hline
\multicolumn{5}{|c|}{${\cal W}={\cal O}^{d=6}_i/M^3_X$} \\
\hline
& Operators ${\cal O}_i$ & $B$& $L$ & $B-L$ \\
\hline
 1&$(\epsilon \dbar_{A_1}\dbar_{B_1}\ubar_{C_1})
(\epsilon \dbar_{A_2}\dbar_{B_2}\ubar_{C_2})$ & $-2$ & $~~0$ & $-2$ \\
\hline
2&$(\epsilon \dbar_{A_1}\dbar_{B_1}\ubar_{C_1})(Q_{A_2}L_{B_2})\dbar_{C_2}$
& $-1$ & $~~1$ & $-2$ \\
\hline
3&$(\epsilon \dbar_{A_1}\dbar_{B_1}\ubar_{C_1})(L_{A_2}L_{B_2})\ebar_{C_2}$ & $-1$
& $~~1$ & $-2$\\
\hline
4&$(\epsilon \dbar_{A}\dbar_{B}\ubar_{C})(\bar{H}\bar{H})\ebar_D$ & $-1$
& $-1$ & $~~0$\\
\hline
 5& $(Q_AL_B)(Q_CL_D)\dbar_E\dbar_F$ & $~~0$   & $~~2$ & $-2$\\
\hline
 6&$(Q_AL_B)(L_CL_D)\dbar_E\ebar_F$ & $~~0$   & $~~2$ & $-2$\\
\hline
7&$(Q_AL_B)(\bar{H}\bar{H})\dbar_C\ebar_D$ & $~~0$ & $~~0$ & $~~0$\\
\hline
8&$(L_AL_B)(L_CL_D)\ebar_E\ebar_F$ & $~~0$   & $~~2$ & $-2$ \\
\hline
9&$(L_AL_B)(\bar{H}\bar{H})\ebar_C\ebar_D$ & $~~0$   & $~~0$ & $~~0$ \\
\hline
10&$(\bar{H}\bar{H})(\bar{H}\bar{H})\ebar_A\ebar_B$ & $~~0$   & $-2$ & $~~2$ \\
\hline
11&$(L_AH)(L_BH)(\bar{H}H)$ & $~~0$ & $~~2$ & $-2$ \\
\hline
12&$(Q_A\bar{H})(L_BH)\ubar_C\ebar_D$ & $~~0$ & $~~0$ & $~~0$ \\
\hline
13&$\epsilon (Q_AQ_B)(Q_C\bar{H})(L_DH)$ & $~~1$ & $~~1$ & $~~0$ \\
\hline
\multirow{2}{*}{14} & $(L_AH)(L_B\bar{H})(HH)$, &\multirow{2}{*}{$0$}
&\multirow{2}{*}{$2$}&\multirow{2}{*}{$-2$} \\ &  $(L_AH)(L_BH)(H\bar{H})$ &&&\\
% \multirow{2}{*}{$0$}
%14&$(L_AH)[(L_B\bar{H})(HH)/(L_BH)(H\bar{H})]$ & $~~0$ & $~~2$ & $-2$ \\
\hline
\end{tabular}
\caption{Sextic Effective  Superpotential Operators from
NMSGUT.}\label{tab:dim7}
\end{center}
\end{table}

\renewcommand{\arraystretch}{1.5}
\begin{table}[h!]
\begin{center}
\begin{tabular}{|p{.4cm}|p{3.7cm}|p{.5cm}|p{.5cm}|p{0.8cm}|}
\hline
\multicolumn{5}{|c|}{${\cal W}={\cal O}^{d=6}_i/M^3_X$} \\
\hline
& Operators ${\cal O}_i$ & $B$& $L$ & $B-L$ \\
\hline % Commented operators are repeated
%1&$(h\bar{h})\epsilon Q_A Q_B Q_C L_D$ & $~~1$ & $~~1$ & $~~0$ \\
1&$(H\bar{H})Q_AL_B\ubar_C\ebar_D$ & $~~0$   & $~~0$ & $~~0$\\
\hline
%3&$(\bar{h}\bar{h})Q_AL_B\bar{d}_C\bar{e}_D$ & $~~0$   & $~~0$ & $~~0$\\
2&$(H\bar{H})Q_AQ_B\ubar_C\dbar_D$ & $~~0$   & $~~0$ & $~~0$\\
\hline
3&$(HH)Q_AQ_B\ubar_C\ubar_D$ & $~~0$   & $~~0$ & $~~0$\\
\hline
4&$(\bar{H}\bar{H})Q_AQ_B\dbar_C\dbar_D$ & $~~0$   & $~~0$ & $~~0$\\
\hline
5&$(H\bar{H})\epsilon \ebar_A \ubar_B \ubar_C \dbar_D$ & $-1$ & $-1$ & $~~0$ \\
\hline
%8&$(\bar{h}\bar{h})\epsilon \bar{e}_A \bar{u}_B \bar{d}_C \bar{d}_D$ & $-1$ & $-1$ & $~~0$ \\
%9&$(h\bar{h})[(L_AL_B)(h_1h_1)/(L_Ah_1)(L_Bh_1)]$ & $~~0$ & $~~2$ & $-2$ \\
\multirow{2}{*}{6} & $(H\bar{H})(\bar{H}H)(\bar{H}H)$ &\multirow{2}{*}{$0$}
&\multirow{2}{*}{$0$}&\multirow{2}{*}{$0$} \\ & $(H\bar{H})(HH)(\bar{H}\bar{H})$ &&&\\
% \multirow{2}{*}{$0$}
%& \multirow{2}{*}{$0$} & \multirow{2}{*}{$0$} \\
\hline
\end{tabular}
\caption{Sextic Effective  Superpotential Operators from ${\cal W}^{GRB}$ part of NMSGUT.}\label{tab:dim7a}
\end{center}
\end{table}

\section*{Acknowledgement}
CSA acknowledges DST research grant EMR/2014/000250.
SK acknowledges DST research grant EMR/2014/000250 for post doctoral support.
RLA acknowledges SERB-DST for national postdoctoral
fellowship, PDF/2016/000863.

\raggedright

\end{document}